\documentclass[prc,aps,groupedaddress,superscriptaddress,twocolumn,nofootinbib,showpacs,showkeys,floatfix,a4paper,10pt,amsmath,amssymb]{revtex4-1}
\usepackage{graphicx}
\usepackage{dcolumn}
\usepackage{bm}

\begin{document}

\title{Microscopic description of spontaneous fission  based on a Gogny energy density 
functional including tensor contributions}

\author{R. Rodr\'{\i}guez-Guzm\'an}
\email{raynerrobertorodriguez@gmail.com}
\affiliation{Physics Department, Nazarbayev 
University, 53 Kabanbay Batyr Ave., Astana 010000, Kazakhstan}

\author{L. M. Robledo}
\email{luis.robledo@uam.es}
\affiliation{%
Center for Computational Simulation, Universidad Polit\'ecnica de 
Madrid, Campus Montegancedo, 28660 Boadilla del Monte, Madrid, Spain
}%
\affiliation{Departamento  de F\'{\i}sica Te\'orica and CIAFF, 
Universidad Aut\'onoma de Madrid, 28049-Madrid, Spain}

\author{R. N. Bernard}
\email{remi.bernard@cea.fr}
\affiliation{CEA, DES, IRESNE, DER, SPRC, LEPh, 13115
Saint-Paul-lès-Durance, France}

\date{\today}

\begin{abstract} This paper extends previous studies on the impact of 
tensor forces in fission dynamics of neutron-deficient Thorium isotopes 
to other isotopic chains of heavy actinides and low-mass super-heavy nuclei. 
Calculations are carried out within a mean-field framework based on the Gogny-D1S 
parametrization supplemented with the D1ST2a perturbative tensor term 
as driving force. Fission barrier heights and spontaneous fission 
half-lives are used as benchmarks to analyze the impact of the tensor 
term. A significant reduction of fission barrier heights and half-lives
is associated to the tensor component of the force.
\end{abstract}


\maketitle{}

%
%
%

\section{Introduction}

Nuclear fission is the consequence of large-amplitude collective motion 
driving the initial configuration (mostly the ground state of the 
parent nucleus) to scission point. Dynamics strongly depends on the 
nuclear interaction used as the associated (quantum) shell effects and 
pairing properties are responsible for tunneling through different 
stages of the evolution to scission. In addition, many-body methods 
used to (approximately) describe fission must be rooted in quantum 
mechanics for a sound microscopic description of the phenomenon. In 
spite of the many years passed since the experimental discovery of 
fission \cite{Hahn:1939} and its initial interpretation \cite{Meitner:1939,Bohr:1939}, the nuclear physics community 
has not reached a consensus on the optimal phenomenological interaction 
to be used to describe fission \cite{Bender:2020}. The consensus is necessary as the 
so-called {\it ab-initio} interactions inspired in the fundamental 
theory of strong interactions are still very far away to provide a 
consistent description of the phenomenon. There is also a vivid debate 
on the most suitable many-body method to be used to describe fission 
observables like spontaneous fission life-time or fission fragment mass 
distribution. Among the wide variety of approaches employed to describe 
fission  \cite{ref1}, the constrained mean-field framework \cite{ref2} 
has already been shown to represent a valuable tool, complementary to 
other theoretical models, to be used as zero order approximation to the 
problem. 

Once the phenomenological interaction is chosen, fission is understood 
at the mean-field level in terms of a fission path parametrized as a 
function of the evolution of deformation parameters characterizing the 
different shapes involved. Associated to the fission path one considers 
a potential energy, obtained at the mean-field level plus some 
corrections to be defined latter, with an associated complex landscape 
-comprising minima, valleys, ridges and saddle points consequence of 
shell effects. Given the importance of pairing, the potential energy is  
built using the Hartree-Fock-Bogoliubov  (HFB) method with constrains 
on several deformation parameters denoted in the following as 
${\bf{Q}}$ \cite{ref1,ref3}. In addition to  energies 
$E_{HFB}({\bf{Q}})$, a microscopic description of fission requires  
collective inertias for the deformation degrees of freedom as well as 
the quantum zero-point vibrational $E_{VIB}({\bf{Q}})$ corrections. The 
fission path involves strongly deformed configurations for which the 
rotational correction to the total energy $E_{ROT}({\bf{Q}})$ is not a 
negligible quantity both because of its magnitude as well as its 
behavior as a function of $\bf{Q}$. All these ingredients define the 
collective potential $V({\bf{Q}}) = E_{HFB}({\bf{Q}})- 
E_{VIB}({\bf{Q}}) - E_{ROT}({\bf{Q}})$ \cite{ref3,ref4,ref5,ref6}. Both 
the collective inertias and the potential $V({\bf{Q}})$, are employed to 
compute the spontaneous fission half-lives $t_\textrm{SF}$ using the 
Wentzel-Kramers-Brillouin (WKB) approximation \cite{ref3}. Those 
collective inertias  and potential $V({\bf{Q}})$ also represent basic 
ingredients within the Time Dependent Generator Coordinate Method 
(TDGCM) in the Gaussian Overlap Approximation (GOA), employed to extract 
observables such as the charge and/or mass distributions of the fission 
fragments \cite{ref8-TDGCM,ref9-TDGCM,ref10-TDGCM}.

Microscopic  fission calculations are usually carried out using the 
least energy (LE) scheme. In this case, for a set of constrains 
${\bf{Q}}$ defining a configuration along the fission path, the HFB 
energy $E_{HFB}({\bf{Q}})$ is minimized according to the 
Ritz-variational principle \cite{ref2}. The quantum zero-point   
corrections are usually included {\it{a posteriori}} \cite{ref3}. Among the assorted 
repertoire of phenomenological interactions, LE 
calculations are typically carried out in terms of  nonreativistic Gogny 
\cite{ref3,ref4,ref5,ref6,ref11,ref12,ref13,ref14,ref15,ref16,refaqui-1,refaqui-2}, 
Skyrme \cite{ref17,ref18,ref19,ref20,ref21} and 
Barcelona-Catania-Paris-Madrid (BCPM) \cite{ref22,ref23,ref24} as well 
as relativistic \cite{ref25,ref26,ref27,ref28,ref29,ref30,ref31} energy 
density functionals (EDFs).

A second class of calculations -which have already received close 
scrutiny using both non-relativistic and relativistic approaches- 
resorts to the least action (LA) scheme. Here, the quantum zero-point  
corrections are also included {\it{a posteriori}} \cite{ref42}. A key 
result, emerging from previous LA studies, is that the  collective 
action gets strongly quenched when pairing degrees of freedom are 
included, leading to a substantial decrease  of several orders of 
magnitude in the computed $t_\textrm{SF}$ values. As a consequence, a 
much better agreement with the available experimental data 
\cite{ref36,ref37,other-MA_Ra,ref38,ref39,ref40,ref-MA_Skyrme,ref41,ref42} 
is obtained in comparison with LE calculations. 
    
Regardless of the LE and/or LA scheme considered, an implicit  
assumption in microscopic fission calculations is that, fission-related 
properties are determined by general features of the effective 
interactions employed. However, interactions are usually tuned to 
reproduce nuclear matter parameters -such as the surface energy of 
semi-infinite nuclear matter- that are not properly constrained by 
experimental data and it is important to include fission data in their 
fitting protocol as done in Refs. 
\cite{ref11,D1S-other,JPG-review,ref17,UNEDF-other}.

In the case of the Gogny-EDF, the parametrization D1S was introduced to 
reduce the too large fission barrier height obtained with the previous 
parametrization and  subsequently applied to the microscopic 
description of fission in heavy and superheavy nuclei 
\cite{ref12,ref15,ref16}. This parametrization, considered a well 
tested standard within the Gogny-D1 family, has shown its ability to 
reproduce reasonably well several other  properties all over the 
nuclear chart both at the mean-field level and beyond 
\cite{JPG-review}. Nevertheless, it has also been found that the 
Gogny-D1S EDF is not specially good in reproducing masses when moving 
away from the stability valley. To cure this deficiency, new 
parametrizations -such as D1N \cite{gogny-d1n} and D1M 
\cite{gogny-d1m}- have been introduced in the Gogny-D1 family. Previous 
studies have already shown that, while improving the description of 
nuclear masses, D1M  essentially retains the same predictive power as 
D1S  to account for fission properties in heavy and superheavy nuclei 
\cite{ref3,ref4,ref5,ref6,refaqui-1,refaqui-2,other-MA_Ra,ref42} as 
well as for a wealth of low-energy nuclear structure data 
\cite{PRCQ2Q3-2012,PTpaper-Rayner,Rayner-Robledo-JPG-2009,Robledo-Rayner-JPG-2012,Rayner-PRC-2010,Rayner-PRC-2011,JPG-review,Dy-Q2Q3,SHV-Q2Q3}.

In spite of their reasonable performance, none of the members of the 
Gogny-D1 family already mentioned  include a tensor force. Such tensor 
components have  received  attention in Skyrme mean-field calculations 
for both spherical and deformed nuclei 
\cite{Tensor-Skyrme-1,Tensor-Skyrme-2}. In the case of the Gogny-EDF, 
like-particle tensor components were first considered to improve the 
evolution of spherical single-particle states along isotopic chains, 
but no attention was  paid to pairing correlations 
\cite{Tensor_Gogny-Otsuka}. Later, a complete long range tensor has 
been added perturbatively to the standard Gogny-D1S EDF 
\cite{Marta-tensor-1,Marta-tensor-2,Marta-tensor-3,Marta-tensor-4,Marta-tensor-5}. 
The reshuffling of single particle orbitals around the Fermi level 
brought about by the tensor interaction can have a sizable impact both 
on the amount of shell effects responsible for barrier properties as 
well as in the strength of pairing correlations crucial to determine 
collective inertias.

Previous studies \cite{OTEM-0,OTEM-1,OTEM-2} have shown that 
second-order tensor corrections mostly affect the (S = 1, T = 0) 
channel and, in the fitting protocol of the original Gogny-EDF, most of 
the effect of the tensor force was already phenomenologically included 
in the strength of the density-dependent term, which also acts in the 
(S = 1, T = 0) channel. Therefore, the inclusion of a  perturbative 
long range tensor in the parametrization D1ST2a (to be referred simply 
as D1ST hereafter) 
\cite{Marta-tensor-1,Marta-tensor-2,Marta-tensor-3,Marta-tensor-4,Marta-tensor-5} 
represents a meaningful first step towards a full refitting of all the 
parameters of the Gogny-EDF. 

Much work still remains to be done to  asses the impact of a tensor 
term like the one in  Gogny-D1ST in the microscopic description of 
fission properties in heavy and superheavy nuclei. The first 
exploration of the role of tensor contributions in fission properties 
was carried out on a fission calculation for a selected set of neutron 
deficient Th isotopes  \cite{Fission-tensor-2020} for which recent 
experimental data has been obtained \cite{mode-act} in the SOFIA collaboration \cite{SOFIA}. It is found, that 
tensor contributions are crucial for the  opening a new  valley, 
associated with the exotic symmetric bimodal fission observed in those 
neutron-deficient Th isotopes \cite{mode-act,mode-act2}.

To the best of our knowledge, a survey of spontaneous fission 
half-lives and related properties, using the Gogny-D1ST EDF, has not 
yet been presented in the literature. This is precisely, the goal of 
this mean-field study using the LE scheme. By using the D1ST 
parametrization along with the HFB approximation \cite{ref2} 
calculations for the isotopic chains $^{240-250}$Cm and $^{240-250}$Cf, 
for which experimental data are available \cite{tsf-exp}, are carried 
out and the outcome compared with results obtained with the standard  
D1S parametrization. Comparison of the predicted lifetimes and other 
properties, obtained with both EDFs, represents a timely and necessary 
step to better understand the role of tensor contributions in the  
fission properties of those isotopes belonging to a region of the 
nuclear chart where several key features of the shell effects 
associated with superheavy elements start to manifest 
\cite{ref3,ref4,refaqui-1,ref42}. Moreover, we have  extended the 
calculations to the low-Z superheavy nuclei $^{242-258}$Fm. Such a 
chain, with a belt-shaped dependence of the spontaneous fission 
half-lives as functions of the mass number \cite{tsf-exp}, represents a 
challenging benchmark to further examine the fission properties 
associated to the Gogny-D1ST EDF.

The paper is organized as follows. In Sec.~\ref{Theory}, we briefly 
outline the theoretical LE framework  employed to study the fission 
properties of the considered nuclei. The results of our calculations 
are discussed in Sec.~\ref{results}. We start the section, with a 
description of our D1S and D1ST Gogny-HFB calculations for the nucleus 
$^{248}$Cf taken as an illustrative example. The same methodology has 
been employed for all the considered nuclei. Next, we present the D1S 
and D1ST systematic of the fission paths and spontaneous fission 
half-lives for the studied Curium  and Californium isotopes. The 
section ends, with a discussion of the D1S and D1ST systematic of the 
spontaneous fission half-lives obtained for $^{242-258}$Fm. In 
Sec.~\ref{results}, we also compare with the available experimental 
$t_{SF}$ values \cite{tsf-exp}.  Finally, Sec.~\ref{conclusions} is 
devoted to the concluding remarks.

%

\section{Theoretical framework}
\label{Theory}
The 
results discussed in this paper, have been obtained with the 
standard D1S \cite{ref11} and D1ST \cite{Marta-tensor-1,Marta-tensor-2,Marta-tensor-3,Marta-tensor-4,Marta-tensor-5}
parametrization of the Gogny-EDF. The effective 
two-body D1ST force reads
\begin{eqnarray} \label{force_D1ST}
V(\vec{r})_{\textbf D1ST} &=& \sum_{i=1,2}
\Big(W_{i} + B_{i}P_{12}^{\sigma}+ H_{i}P_{12}^{\tau} + W_{i}P_{12}^{\sigma}P_{12}^{\tau}
\Big) e^{-\frac{\vec{r}^{2}}{\mu_{i}^{2}}}
\nonumber\\
&+&
t_{0} \Big(1+ x_{0}P_{12}^{\sigma}\Big) \rho^{\alpha}(\vec{R}) \delta(\vec{r})
\nonumber\\
&+&
W_{LS} \overleftarrow{\nabla} \delta(\vec{r}) \wedge \overrightarrow{\nabla}
\cdot ~ \Big(\vec{\sigma}_{1} \cdot \vec{\sigma}_{2} \Big)
\nonumber\\
&+&
\Big(V_{T1}+ V_{T2} P_{12}^{\tau}\Big) S_{12}(\vec{r})
e^{-\frac{\vec{r}^{2}}{\mu_{TS}^{2}}}
\end{eqnarray}
where
\begin{eqnarray}
S_{12}(\vec{r}) = 3 ~
\frac{\Big(\vec{\sigma}_{1} \cdot \vec{r} \Big) \Big(\vec{\sigma}_{2} \cdot \vec{r} \Big)}{r^{2}}
- \vec{\sigma}_{1} \cdot \vec{\sigma}_{2}
\end{eqnarray}
In this expression $\vec{r}$ and $\vec{R}$ stand for the relative and center of mass
coordinate of two particles.

\begin{figure*} 
\includegraphics[width=0.9\textwidth]{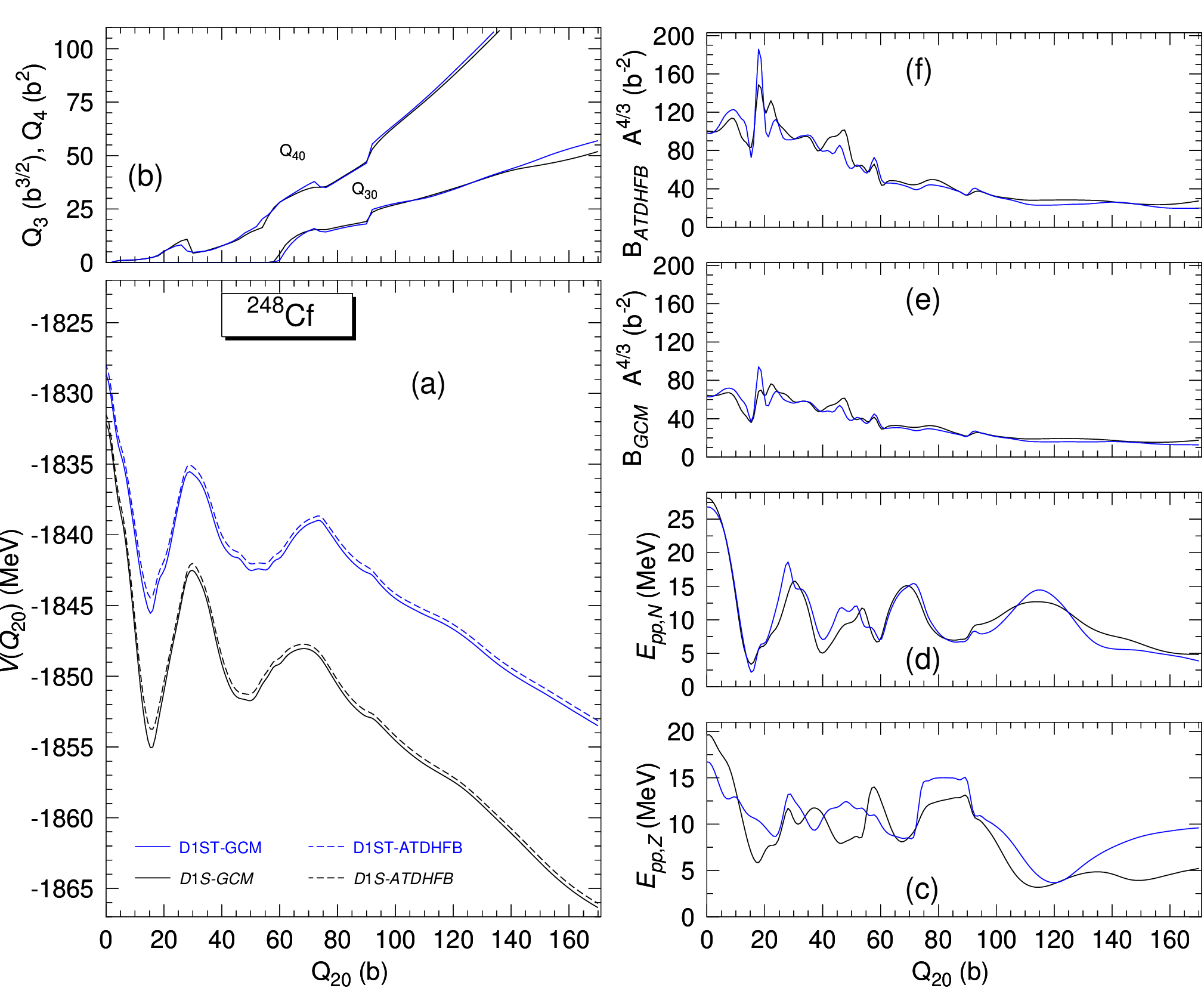}
\caption{(Color online) 
The collective potentials $V(Q_{20})$  Eq.(\ref{Coll-Pot-V-Q20}) 
obtained for the nucleus $^{248}$Cf, within the GCM and ATDHFB schemes, 
are plotted in panel (a) as functions of the quadrupole moment 
$Q_{20}$. The octupole $Q_{30}$ and hexadecapole $Q_{40}$ moments of 
the intrinsic states are plotted in panel (b). The proton $E_{pp,Z}$ 
and neutron $E_{pp,N}$ pairing interaction energies are depicted in 
panels (c) and (d), while the collective GCM and ATDHFB masses are 
plotted in panels (e) and (f). Results have been obtained with the 
parametrizations D1S and D1ST of the Gogny-EDF. For more details, see 
the main text.
}
\label{example} 
\end{figure*}

The first three lines in Eq.(\ref{force_D1ST}) correspond to any 
standard member of the Gogny-D1 family, the only differences being in 
the values of the parameters. In the case of the effective D1ST force, 
the values of the parameters $\Big\{W_{i} , B_{i} , H_{i} , M_{i}, 
\mu_{i}, i = 1, 2 \Big\}$, $t_{0}$, $x_{0}$, $\alpha$ and $W_{LS}$ are 
the same as the ones in the standard Gogny-D1S force \cite{ref11}. On 
the other hand, the tensor parameters $V_{T1}$ and $V_{T2}$ have been  
adjusted to reproduce the $1 f_{5/2}$ and $1 f_{ 7/2}$  neutron 
single-particle energies in $^{48}$Ca. The range of the spatial form 
factor is taken equal to the largest range of the central potential of 
D1S, namely $\mu_{TS} =$ 1.2 fm. 

Axial and simplex symmetries have been kept as self-consistent symmetries \cite{ref2}. 
In addition to the usual HFB constrains on the proton and neutron 
number operators \cite{ref2}, we have employed constrains on the 
quadrupole $\hat{Q}_{20} = z^{2} - \frac{1}{2}\Big (x^{2} + y^{2})z$ 
and octupole $\hat{Q}_{30} = z^{3} -\frac{3}{2}\Big (x^{2} + y^{2})z $ 
operators \cite{PRCQ2Q3-2012,Robledo-Rayner-JPG-2012,ref42}
to obtain the D1S and/or D1ST  deformation properties.

The quasiparticle operators \cite{ref2} have been expanded
in a deformed axially symmetric harmonic oscillator (HO)
basis containing states with $J_{z}$ quantum numbers up to
35/2 and up to 26 quanta in the z-direction.  
For each of the 
configurations along the fission paths of the studied nuclei, the 
HO lengths b$_{\perp}$ and b$_{z}$ have been optimized so as
to minimize the total HFB energy.

For the solution of the HFB equations, an approximate second order 
gradient method \cite{ref60} has been used. The two-body kinetic energy 
correction has been fully taken into account in the Ritz-variational 
\cite{ref2} procedure. In the calculations, Coulomb exchange is 
evaluated in the Slater approximation \cite{ref61} while Coulomb and 
spin-orbit antipairing are neglected. As in previous works 
\cite{Marta-tensor-1,Marta-tensor-2,Marta-tensor-3,Marta-tensor-4,Marta-tensor-5,Fission-tensor-2020}, 
the tensor contribution is only included in the Hartree-Fock field 
while its contribution to the pairing field is neglected. Preliminary results
seem to indicate that the tensor pairing field is small as compared to
other contributions and has little impact on fission dynamics.

We have computed the spontaneous fission half-lives $t_\textrm{SF}$ (in 
seconds) using the WKB formalism
\begin{equation} \label{TSF}
t_\mathrm{SF}= 2.86 \times 10^{-21} \times \left(1+ e^{2S} \right)
\end{equation}
where the action $S$ along the (one-dimensional $Q_{20}$-projected)
fission path reads
\begin{equation} \label{Action}
S= \int_{a}^{b} dQ_{20} S(Q_{20})
\end{equation}
with the integrand 
\begin{equation}
\label{integrand-eq}
S(Q_{20}) = \sqrt{2B(Q_{20})\left(V(Q_{20})-\left(E_{Min}+E_{0} \right)  \right)}
\end{equation}
The integration limits $a$ and $b$ in Eq.(\ref{Action}) correspond to  
classical turning points \cite{turning} evaluated for the energy 
$E_{Min} + E_{0}$. The energy $E_{Min}$ corresponds to the absolute 
minimum of the considered path, while $E_{0}$ accounts for the true 
ground state energy once quadrupole fluctuations are taken into 
account. This quantity can be estimated with a harmonic 
approximation using the curvature of the potential around the minimum 
and the collective inertia of the quadrupole motion. Typical values 
obtained in the region are around 0.5 MeV with some fluctuations 
associated to variations with neutron number. In order to better 
disentangle the effect of the tensor term a constant value $E_{0} =$ 
0.5 MeV has been employed in this work.

In Eq.(\ref{integrand-eq}), $B(Q_{20})$ represents the quadrupole collective mass, while  
the collective potential $V(Q_{20})$ reads 
\cite{ref3,ref4,ref5,ref6}
\begin{eqnarray} \label{Coll-Pot-V-Q20}
V(Q_{20}) = E_{HFB}(Q_{20})- E_{VIB}(Q_{20}) 
- E_{ROT}(Q_{20}) 
\end{eqnarray}
with $E_{HFB}(Q_{20})$, $E_{VIB}(Q_{20})$ and $E_{ROT}(Q_{20})$ being the 
HFB, vibrational and rotational correction energies, respectively, as functions 
of the quadrupole deformation  $Q_{20}$.

The collective mass $B(Q_{20})$ as well as the zero-point vibrational 
energy $E_{VIB}(Q_{20})$ have been computed using the perturbative 
cranking approximation \cite{ref3,ref65,ref66,ref67} to the Adiabatic 
Time Dependent HFB (ATDHFB) and  the Gaussian Overlap Approximation 
(GOA) \cite{ref3,ref2}. The zero-point rotational energy 
$E_{ROT}(Q_{20})$ has been computed in terms of the Yoccoz moment of 
inertia \cite{ref68,ref69,ref70}. Therefore, in this study two 
different paths -the GCM and the ATDHFB paths- will be considered for 
each nucleus with the D1S and D1ST EDFs.

%


\section{Discussion of the results}
\label{results}


In this section, the results obtained for the $^{240-250}$Cm, 
$^{240-250}$Cf and $^{242-258}$Fm isotopic chains are discussed. Let us 
first describe in more detail the results obtained for the nucleus 
$^{248}$Cf, taken as an illustrative example. The GCM and ATDHFB 
collective potentials $V(Q_{20})$  Eq.(\ref{Coll-Pot-V-Q20}), obtained 
for this nucleus, are plotted in panel (a) of Fig.~\ref{example} as 
functions of the quadrupole moment $Q_{20}$ of the intrinsic states.

The first noticeable feature, is the  underbinding of the D1ST curves 
with respect  to the D1S ones. For example, for $Q_{20}$ = 16~b, 
$\delta V_{GCM} = V_{GCM}^{D1ST}- V_{GCM}^{D1S} =$ 9.64 MeV and $\delta 
V_{ATDHFB} = V_{ATDHFB}^{D1ST}- V_{ATDHFB}^{D1S} =$ 9.39 MeV, 
respectively. Such a global underbinding of the D1ST HFB solutions 
-arising from the proton-neutron part of the tensor term- is a common 
feature for all the nuclei studied in this work and is a consequence of 
the features of the tensor term considered. Nevertheless, 
it is satisfying to observe that all the  curves in panel (a) look 
rather  similar with the ground state located at $Q_{20}$ = 16~b and a 
fission isomeric state at $Q_{20}$ = 50~b. However, although the fission 
paths look quantitatively the same they differ in their barrier 
height. For D1S  the top of the inner barrier is located around 
$Q_{20}$ = 30~b, and its height reaches the values $B_{I,GCM} =$ 12.55 
MeV, $B_{I,ATDHFB} =$ 11.74 MeV whereas for D1ST one obtains $B_{I,GCM} 
=$ 9.79 MeV and $B_{I,ATDHFB} =$ 9.20 MeV, respectively, which are 
significantly lower than the D1S results. One has to keep in mind that 
triaxiality is not allowed in the calculation, a restriction that 
increases the barrier heights by one or two MeV. This effect cannot be 
included at the present stage along with the tensor force.

The 
reflection-symmetric fission isomer corresponding to the second excited 
minimum is located at $Q_{20}=$ 52~b in the D1S case while it is shifted 
to a slightly larger quadrupole moment when the tensor interaction is 
considered. Excitation energies are $E_{II,GCM}^{D1S} =$ 3.34 MeV, 
$E_{II,ATDHFB}^{D1S} =$ 2.52 MeV, $E_{II,GCM}^{D1ST} =$ 2.90 MeV and 
$E_{II,ATDHFB}^{D1ST} =$ 2.34 MeV. The effect of the tensor is to 
slightly reduce the fission isomer excitation energy but the effect is, 
by no means, relevant for fission dynamics. Regardless of the Gogny-EDF 
employed, octupole correlations play a prominent role in the outer 
sectors of the GCM and ATDHFB paths and are responsible for the 
asymmetric fission fragment mass distribution observed experimentally. 
In addition, octupole correlations significantly reduce the height 
$B_{II}$ of the outer fission barrier. The position of the second 
barriers is predicted at 70~b (D1S) and 76~b (D1ST). The height of the 
second barrier obtained with D1S are  $B_{II,GCM} =$ 7.01 MeV and 
$B_{II,ATDHFB} =$ 6.04 MeV, whereas sligthly lower values are obtained 
for D1ST $B_{II,GCM} =$ 6.43 MeV and $B_{II,ATDHFB} =$ 5.70 MeV. 

\begin{figure*}
\includegraphics[width=0.80\textwidth]{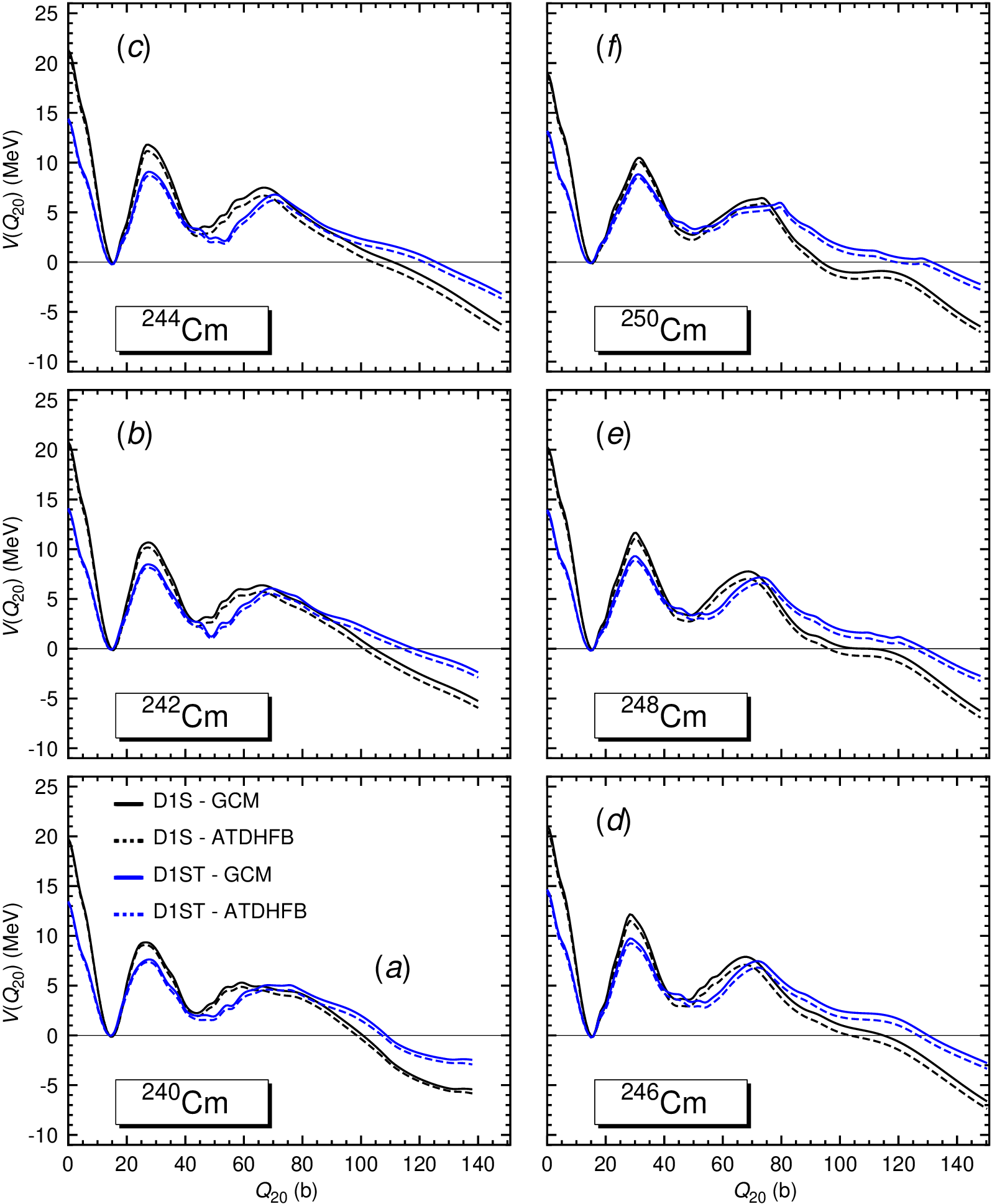}
\caption{(Color online) The collective potentials $V(Q_{20})$ 
Eq.(\ref{Coll-Pot-V-Q20}) obtained for the nuclei $^{240-250}$Cm, 
within the GCM and ATDHFB schemes, are plotted as functions of the 
quadrupole moment $Q_{20}$. All the relative energies are measured with 
respect to the absolute minima of the corresponding paths. Results have 
been obtained with the parametrizations D1S and D1ST of the Gogny-EDF. 
For more details, see the main text.
}
\label{FissionPAths_Cm} 
\end{figure*}

One 
can summarize the previous discussion by saying that the main effect of 
the tensor interaction is to reduce significantly the first barrier 
height. It is also responsible for a slight displacement of the 
position of the two minima, and the second barrier to larger values of 
the quadrupole moment. Finally, excitation energies of the fission 
isomer as well as second barrier heights are not modified in a substantial 
way. These effects are also visible in other isotopes, as it will be 
discussed below.
 
\begin{figure*}
\includegraphics[width=0.80\textwidth]{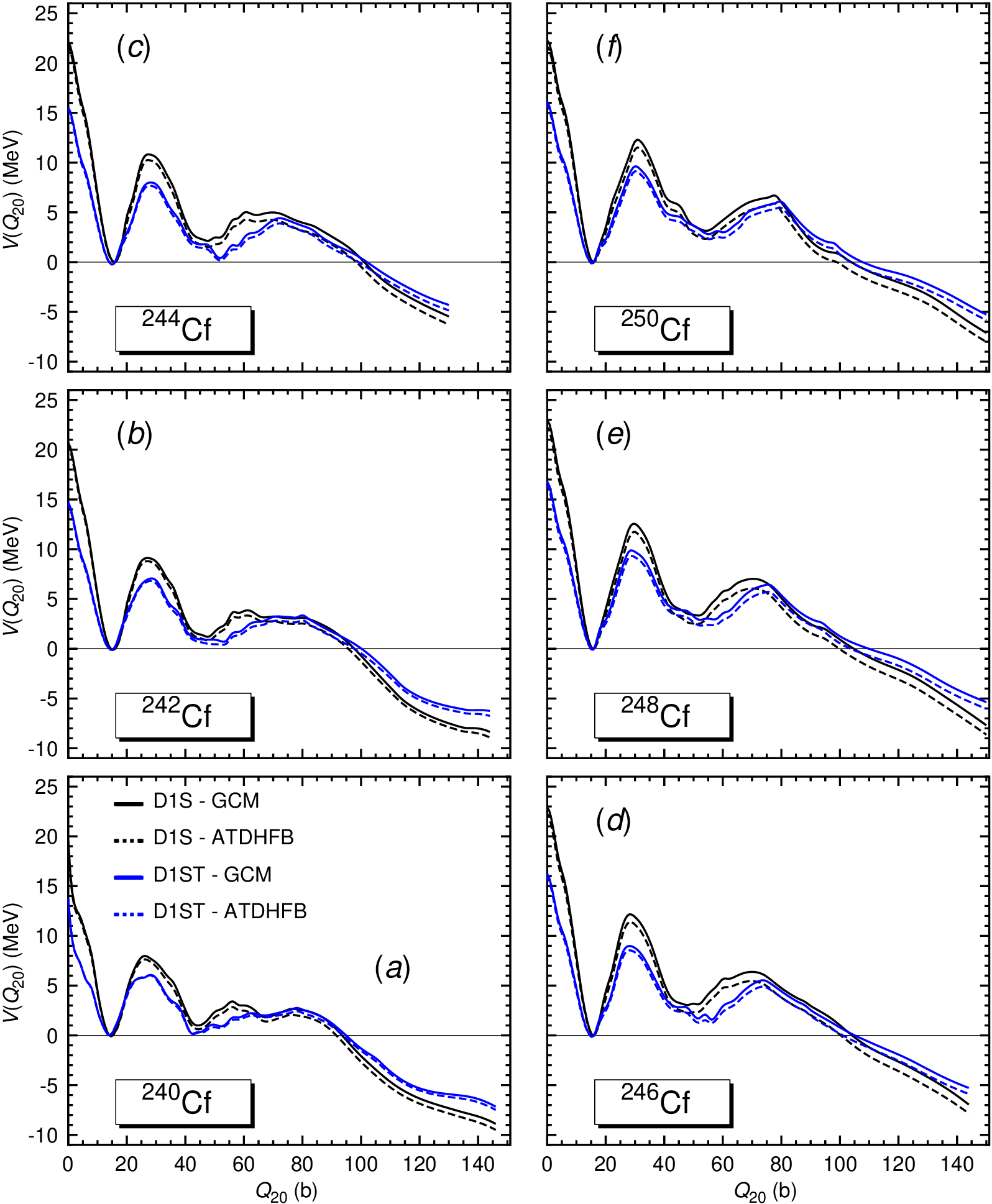}
\caption{(Color online) The same as Fig.~\ref{FissionPAths_Cm} but for the nuclei
$^{240-250}$Cf. 
}
\label{FissionPAths_Cf} 
\end{figure*}

The octupole $Q_{30}$ and hexadecupole $Q_{40}$ moments of the 
intrinsic states are plotted in panel (b) of Fig.~\ref{example}. As the 
spatial reflection symmetry can be broken at any stage of the 
calculations an additional constrain on the center of mass operator 
$\hat{Q}_{10}$ has to be imposed to keep the center of mass fixed at 
the origin \cite{PRCQ2Q3-2012,Robledo-Rayner-JPG-2012,ref42} to avoid 
spurious center of mass motion. As can be seen from the panel, the D1S 
and D1ST $Q_{30}$ and $Q_{40}$ values are rather similar. Note also 
from the panel, that for $^{248}$Cf octupole deformations play a role 
for $Q_{20} \ge$ 56-58~b roughly after the fission isomer minimum.

The proton and neutron pairing interaction energies $E_{pp,\tau} = 
\frac{1}{2} \textrm{Tr}(\Delta_{\tau} \kappa_{\tau} )$ (with 
$\tau=Z,N$) of the intrinsic states are plotted in panels (c) and (d).
In the present case, only the Gaussian central potential contributes to
these quantities. Therefore, the different pairing 
content in both functionals reflects the tensor rearrangement of the 
underlying single-particle spectra. As can be seen from panels (c) and 
(d), though some differences are also observed for the neutron-neutron 
pairing interaction energies, the most prominent differences are 
obtained for the proton-proton ones.

\begin{figure*}
\includegraphics[width=0.8\textwidth]{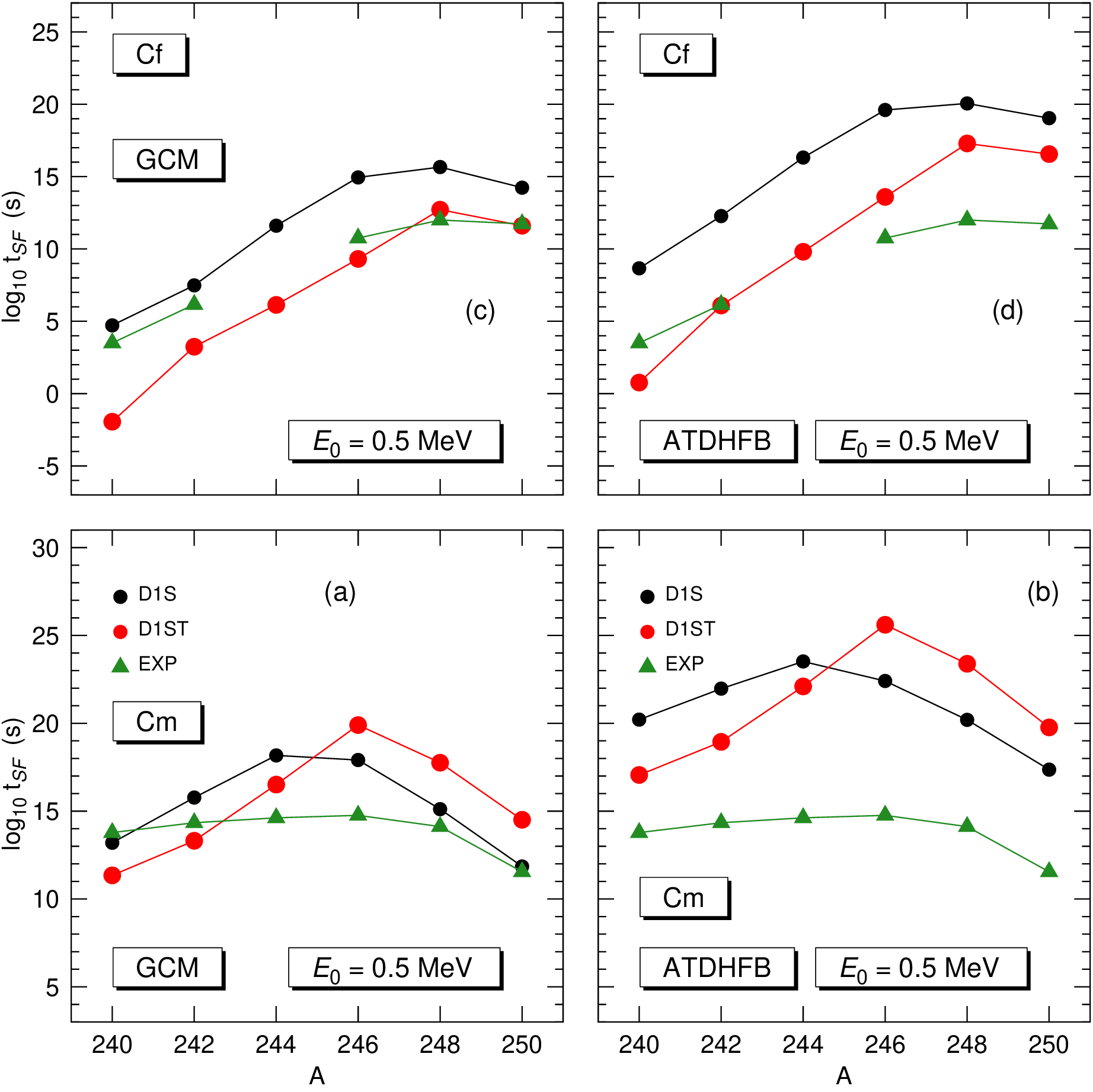}
\caption{(Color online) 
The  spontaneous fission half-lives $t_{SF}$, predicted for the nuclei 
$^{240-250}$Cm [$^{240-250}$Cf], within the GCM and ATDHFB schemes, are 
depicted in panels (a) and (b) [panels (c) and (d)] as functions of the 
mass number. Results have been obtained with the parametrizations D1S 
and D1ST of the Gogny-EDF and $E_{0}= 0.5 ~MeV$. The available 
experimental values, taken from Ref.~\cite{tsf-exp}, are included in 
the plots.
}
\label{LogTSF_CmCf} 
\end{figure*}

The collective GCM and ATDHFB masses $B(Q_{20})$ are depicted in panels 
(e) and (f). As already mentioned, those masses have been computed 
using the cranking approximation, and a three point filter has been 
employed to soften the wiggles associated to level crossings 
\cite{ref3}. For $Q_{20} \le$ 20~b the D1ST masses are slightly larger 
than the D1S ones while for larger quadrupole moments the former are, 
on the average, smaller than the latter. Regardless of the considered 
Gogny-EDF, both the GCM and ATDHFB masses display a similar qualitative 
and quantitative trend. Such a trend is well correlated with the 
patterns observed in the pairing interaction energies in panels (c) and 
(d), and reflects the inverse dependence of the collective masses with 
the square of the pairing gap \cite{turning,inverse-B}. As mentioned 
before, the contribution of the tensor interaction to the pairing 
channel has been neglected. Given the strong dependence of the inertias 
with pairing correlations the pairing associated to the tensor term 
should be studied in depth in the future. Another feature that emerges 
is that both the D1S and D1ST ATDHFB masses are always larger than 
their GCM counterparts. This is the reason to include both kinds of 
collective masses in this study, as their differences can affect the 
predicted lifetimes by several orders of magnitude \cite{ref3,ref4}. In 
the case of $^{248}$Cf and $E_{0}=$ 0.5 MeV, we have obtained the D1S 
(D1ST) GCM and ATDHFB values $log_{10}~t_{SF,GCM}^{D1S}=$ 15.66 
($log_{10}~t_{SF,GCM}^{D1ST}=$ 12.71) and 
$log_{10}~t_{SF,ATDHFB}^{D1S}=$ 20.06 ($log_{10}~t_{SF,ATDHFB}^{D1ST}=$ 
17.29). The larger ATDHFB values are associated to the larger inertia 
that implies a larger action integral. These results should be compared 
with the corresponding experimental value $log_{10}~t_{SF}=$ 12 
\cite{tsf-exp}.

The GCM and ATDHFB collective potentials $V(Q_{20})$, obtained for 
$^{240-250}$Cm and $^{240-250}$Cf are plotted in panels (a) to (f) of 
Figs.~\ref{FissionPAths_Cm} and \ref{FissionPAths_Cf}, as functions of 
$Q_{20}$. All the relative energies are measured with respect to the 
absolute minima of the corresponding paths to facilitate the 
comparisons. Let us remark, however, that for all the considered nuclei 
there is a global underbinding of the D1ST HFB solutions -arising from 
the proton-neutron part of the tensor term- with respect to the D1S 
solutions. In panel (e) of Fig.~\ref{FissionPAths_Cf}, results for  
$^{248}$Cf are included for the sake of completeness.  The excitation 
energies of the spherical configurations predicted with the Gogny-D1ST 
EDF  are always smaller than the corresponding D1S values. For example, 
within the GCM  scheme, the D1ST spherical configuration in $^{244}$Cm 
($^{244}$Cf) is predicted  6.91 (6.56) MeV below the D1S one. For the 
ATDHFB case, the values are very similar to the GCM ones, namely 6.64 
(6.28) MeV.

The absolute minima of the collective potentials correspond to 
$Q_{20}=$ 14-16~b. Inner and outer barriers, as well as 
reflection-symmetric fission isomers in between them, are apparent from 
Figs.~\ref{FissionPAths_Cm} and \ref{FissionPAths_Cf}. As can be seen 
from the figures, smaller barrier heights are obtained with the 
Gogny-D1ST EDF. However, the outer sectors of the fission path obtained 
with this EDF for Cm isotopes, exhibit a much gentler decline than the 
ones obtained with the standard D1S parametrization. This means that 
the tensor contribution is somehow affecting the surface tension 
coefficient of the interaction. However, it has to be determined 
whether a full refit of the force including the tensor term will 
reabsorb this effect into other components of the force. On the other 
hand, such an effect is less pronounced for Cf nuclei and therefore it 
is dependent on mass number.

The summit of the inner barriers is located at $Q_{20}=$ 26-30~b. 
Regardless of the GCM and/or ADTHFB scheme employed, the heights 
$B_{I}$ of the inner barriers in Cm (Cf) isotopes increase as functions 
of the mass number, reaching a maximum at  A = 246 (248). As can be 
seen from the figures, the $B_{I}$ values obtained with the 
parametrization D1ST are smaller than the D1S heights. 

In the case of the outer barriers, their summit is located around 
$Q_{20}=$ 60-74~b and $Q_{20}=$ 70-78~b for the Cm and Cf isotopes, 
respectively. The outer barriers for  $^{240,242}$Cf, display a 
two-humped structure with  the summit located  around $Q_{20}=$ 56~b. 
The summit of the outer barriers is located at slightly larger values 
of the quadrupole moment in calculations carried out with the 
Gogny-D1ST EDF. For the nuclei shown in Figs.~\ref{FissionPAths_Cm} and 
\ref{FissionPAths_Cf}, octupole correlations  play a role for $Q_{20} 
\ge$ 54~b, and affect the height $B_{II}$ of the outer barriers 
significantly. Similar to the inner barriers, the heights $B_{II}$ 
increase as functions of the mass number, reaching a maximum at  A = 
246 (248). 

\begin{figure*}
\includegraphics[width=0.9\textwidth]{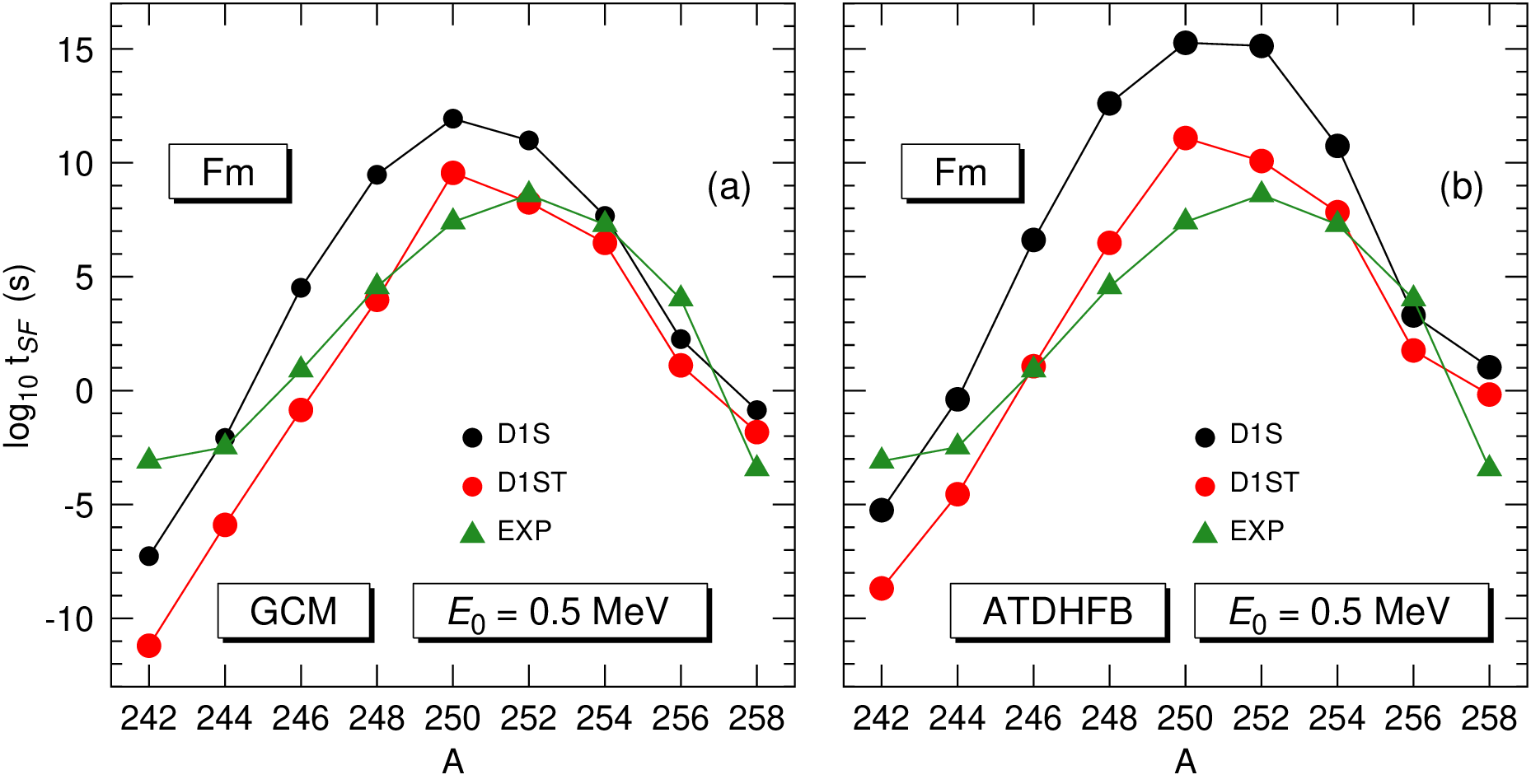}
\caption{(Color online) 
The  spontaneous fission half-lives $t_{SF}$, predicted for the nuclei
$^{242-258}$Fm, within the GCM and ATDHFB schemes, are depicted 
in panels (a) and (b) as functions of the mass number. 
 Results have been obtained with the parametrizations D1S and D1ST of 
the Gogny-EDF and $E_{0}= 0.5 ~MeV$. The available experimental 
values, taken from Ref.~\cite{tsf-exp}, are included in the plots.
}
\label{LogTSF_Fm} 
\end{figure*}

In the calculations,  the smallest $B_{I}$ and $B_{II}$ heights  are 
always predicted within the ATDHFB scheme and with the Gogny-D1ST EDF. 
Even when tensor correlations tend to reduced the heights of the inner 
barriers, the comparison (without triaxiality included) with the 
available experimental $B_{I}^{exp}$ values for $^{242-248}$Cm still 
reveals differences of 1.46, 2.51, 3.21 and 3.41 MeV. For 
$^{242-248}$Cm and $^{250}$Cf the differences between the predicted and 
experimental $B_{II}^{exp}$ values are still 0.52, 1.11, 2.01, 1.77 MeV 
and 1.95 MeV, respectively 
\cite{Refs-barriers-other-nuclei-1,Refs-barriers-other-nuclei-2}. The 
Slater approximation to Coulomb exchange probably plays a role as well 
here: it was shown that the impact of this approximation amounts to up 
to 1 MeV in the Thorium chain \cite{Fission-tensor-2020}.

The previous results suggest that correlations not explicitly taken 
into account in this work might be required to improve the agreement 
with the available experimental data for (static) inner and outer 
barrier heights. For example, previous calculations without tensor 
correlations \cite{ref3}, based on the Gogny-D1M EDF, have found that 
for $^{242-248}$Cm triaxiality reduces the predicted  inner barrier 
height by 1.47, 2.11, 2.72 and  3.18 MeV. At this point, it is worth to 
remark that the heights of (static) fission barriers are not physical 
observables. Their values are inferred from the behavior of fission 
cross sections as a function of the energy and using certain model 
assumptions \cite{not-phys-Naza}. Moreover, for a given nucleus the 
spontaneous fission half-life Eq.~(\ref{TSF}) also depends on the shape 
and width of the barrier and, more generally, via Eqs.~(\ref{Action}) 
and (\ref{integrand-eq}), on the topography of the collective potential 
in between the classical turning points as well as on the collective 
inertias.

A key lesson extracted from previous studies is that, caution must be 
taken in linking barrier heights with  $t_{SF}$ values 
\cite{ref42,ref37,other-MA_Ra}. As can  be seen from Eq. 
(\ref{integrand-eq}), not only the collective potential but also the 
collective mass -with its implicit pairing dependence- does play a role 
in the predicted lifetime. Previous LA studies -in which, the 
collective action is minimized with respect to pairing degrees of 
freedom- have  shown that larger barrier heights do not necessarily  
translate into larger $t_{SF}$ values \cite{ref42,ref37,other-MA_Ra}. 
In fact, it is the subtle interplay between the collective potential 
and the collective mass what leads to the final value for the 
spontaneous fission half-life. Note also, that pairing fluctuations 
within the LA scheme can restore axial symmetry along the fission path 
\cite{ref40,ref-MA_Skyrme} suggesting that the impact of triaxiality in 
the predicted lifetimes is quite limited.

Let us now turn our attention to the spontaneous fission half-lives 
predicted for the nuclei $^{240-250}$Cm and $^{240-250}$Cf with the 
parametrizations D1S and D1ST of the Gogny-EDF. For $^{240-250}$Cm 
[$^{240-250}$Cf] the corresponding GCM and ATDHFB $t_{SF}$ values are 
depicted in panels (a) and (b)  [(c) and (d)] of 
Fig.~\ref{LogTSF_CmCf}, as functions of the mass number. The available 
experimental lifetimes \cite{tsf-exp} are also included in the panels. 
The results shown in the figure, have been obtained with $E_{0} =$ 0.5 
MeV \cite{ref3}. However, we have checked that other values of the 
parameter $E_{0}$ -for example, 1.0 MeV \cite{ref3} and/or the value 
obtained from the curvature around the absolute minimum of the path and 
the collective quadrupole inertia \cite{ref6}- do not alter 
qualitatively any of the conclusions of this work. 

It is satisfying to observe from panels (a) and (c) of 
Fig.~\ref{LogTSF_CmCf} that, for $^{240-250}$Cm and $^{240-250}$Cf, the 
$t_{SF}$ values predicted with both the D1S and D1ST Gogny-EDFs within 
the GCM scheme, reproduce reasonably well the observed experimental 
trend as a  function of the mass number. In the case of $^{240-244}$Cm, 
the spontaneous fission half-lives obtained with the parametrization 
D1ST are smaller than the ones obtained with the standard parameter set 
D1S, while for larger mass numbers the former provides larger values 
than the latter. This behavior can be easily understood by taking into 
consideration first, that the inertias are very similar in the two 
cases and therefore their variations do not lead to substantial changes 
in the lifetimes. Second, but not least, the impact of the tensor term 
in the structure of the fission paths: for $^{240-244}$Cm, the first 
barrier is lower than the one obtained with D1S and the remaining of 
the collective potential looks rather similar. However, for heavier Cm 
isotopes, the much gentler decline of the path in the calculation with 
the tensor term gives an additional contribution to the action in the 
region $Q_{20} > 80 \textrm{b}$.

On the other hand, for the considered Californium isotopes the D1ST 
Gogny-EDF always provides the smallest $t_{SF}$ values as a consequence 
of the lower inner barrier heights and a similar decline of the fission 
path at large deformation. As can be seen from panels (b) and (d) 
similar features are observed for the spontaneous fission half-lives 
predicted within the ATDHFB scheme. This corroborates the robustness of 
the predicted D1S and D1ST trends with respect to the (GCM and/or 
ATDHFB) scheme employed in the calculations. Regardless of the EDF 
employed in the calculations, the ATDHFB $t_{SF}$ values are always 
larger than the corresponding GCM lifetimes as a consequence of the 
larger ATDHFB inertias.

In order to further examine the predictive power of the D1ST 
parametrization, we have extended the calculations to $^{242-258}$Fm. 
The  $t_{SF}$ values, obtained for those Fermium isotopes, within the 
GCM and ATDHFB schemes, are depicted as functions of the mass number in 
panels (a) and (b) of Fig.~\ref{LogTSF_Fm}. Calculations have been 
carried out with  $E_{0}= 0.5 $~MeV. The experimental belt-shaped 
dependence of the spontaneous fission half-lives, as functions of the 
mass number, is well reproduced by the D1S and D1ST Gogny-EDFs. With 
both parameter sets, the largest $t_{SF}$ value is predicted two mass 
units earlier (A=250) than in the experiment. As in the Cm and Cf 
nuclei discussed before, the ATDHFB $t_{SF}$ values are always larger 
than the GCM ones as a consequence of the larger inertias in the former 
case. Moreover, as with the studied Cf isotopes, for $^{242-258}$Fm the 
parametrization D1ST provides smaller spontaneous fission half-lives. 
The explanation is similar to the one offered for the Cf isotopes: the 
first fission barrier is lower with the tensor interaction and the 
gentler decline for large deformation observed in this case is not 
enough as to push the collective potential above $E_{min}+E_{0}$ and 
give a contribution to the action. 

From the results discuss in this section, we conclude that the D1ST 
Gogny-EDF, including tensor contributions perturbatively 
\cite{Marta-tensor-1,Marta-tensor-2,Marta-tensor-3,Marta-tensor-4,Marta-tensor-5},  
represents a reasonable starting point to describe spontaneous fission 
and related properties in heavy and low-Z superheavy nuclei.

\section{Summary and conclusions}
\label{conclusions}

In this paper the impact on fission properties of a perturbative tensor 
term, added to the traditional Gogny D1S force, has been analyzed. 
Fission barrier heights and spontaneous fission half-lives are used as 
benchmark quantities. We observe that the tensor component of the force 
tends to push the position of maxima and minima of the collective 
potential to larger values of the quadrupole moment. The first fission 
barrier height is reduced by a couple of MeV as compared to the 
calculation without tensor. This is also true, but to a lesser extent, 
for the second barrier height. Additionally, the decline of the fission 
path beyond the second fission barrier is gentler when the tensor 
component is included. In the present approach, the tensor contribution 
to the pairing field is neglected and therefore the inclusion of the 
tensor has little impact on the collective inertias. The combination of 
the three above mentioned effects: similar collective inertias, lower 
barrier heights and gentler decline of the collective potential make 
the spontaneous fission half-lives shorter in most of the cases when 
the tensor contribution is included. However, there are exceptions as 
the lower barrier heights and the gentler decline of the path are 
competing effects. From the results obtained in the Cm, Cf and Fm 
isotopic chains we conclude that the perturbative tensor term 
introduced has a non-negligible impact in fission dynamics and should 
be included in the calculations. Within this context, for example, the 
impact of the tensor pairing on the size of collective inertias,  the 
role of triaxiallity as well as the impact of tensor contributions in 
LA computations of spontaneous fission half-lives still deserve a 
detailed study. Work along these lines is in progress and will be 
reported in future publications.

\begin{acknowledgments}
The work of LMR is supported by Spanish Agencia Estatal de Investigacion 
(AEI) of the Ministry of Science and Innovation under Grant No. 
PID2021-127890NB-I00. The work of R. Rodr\'{\i}guez-Guzm\'an was partially 
supported through the grant PID2022-136228NB-C22 funded by 
MCIN/AEI/10.13039/501100011033/FEDER, UE and "ERDF A way of making 
Europe".
\end{acknowledgments}

%
%

\end{document}